\documentstyle[aps,pre,epsfig]{revtex}

\newcommand{\rhoc}{\rho_{\scriptstyle \rm c}}
\newcommand{\tw}{t_{\scriptstyle \rm w}}
\newcommand{\kB}{k_{\scriptscriptstyle \rm B}}
\newcommand{\TEdw}{T_{\scriptscriptstyle \rm Edw}}
\newcommand{\sEdw}{s_{\scriptscriptstyle \rm Edw}}
\newcommand{\Tdyn}{T_{\scriptscriptstyle \rm dyn}}

\begin{document}

\draft

\twocolumn[\hsize\textwidth\columnwidth\hsize\csname@twocolumnfalse\endcsname
    
\title{Effective temperature and compactivity of a lattice-gas under gravity}

\author{Mauro Sellitto}

\address{The Abdus Salam International Centre for Theoretical Physics \\ 
	 Strada Costiera 11, 34100 Trieste, Italy.}

\maketitle

\begin{abstract}
The notion of longitudinal effective temperature and its relation with
the Edwards compactivity are investigated in an abstract lattice gas
model of granular material compacting under gravity and weak thermal
vibration.
\end{abstract}

\pacs{05.40.-a, 05.70.Ln, 67.40.Fd}

\twocolumn\vskip.5pc]\narrowtext

A distinctive feature of mean-field glassy dynamics is a peculiar
violation of fluctuation-dissipation relations which leads to the
definition of a time-scale dependent ``effective
temperature''~\cite{CuKuPe,FrVi}, and the possibility of constructing
a non-equilibrium thermodynamics of glasses and dense granular
media~\cite{BaKuLoSe,Sam}.  Effective temperature also appears in
athermal systems, where the high packing density regime is attained by
compression or by using a confining potential. A particularly
interesting situation, which is relevant to the study of granular
materials, occurs when the confining force is gravity.  In
this case a non stationary inhomogeneous density profile generally
arises and the notion of effective temperature may be then not well
defined unless suitable conditions are verified.  In this note we
explore the possibility of defining a global effective temperature in
an abstract model of granular material under gravity and weak thermal
vibration, and its relation with the Edwards compactivity.

The model consists of a gas of $N$ particles on a body centred cubic
(bcc) lattice where there can be at most one particle per site.  There
is no cohesion energy among particles and the Hamiltonian is
\begin{eqnarray} 
	{\cal H}_0 & = & m g \sum_{i=1}^N h_i \,,
\label{H_0}
\end{eqnarray}
where $g$ is the gravity constant, $h_i$ is the height of the particle
$i$, and $m$ its mass.  At each time step a particle can move with
probability $p$ to a neighboring empty site if the particle has less
than $\nu$ nearest neighbors before and after it has
moved~\cite{KoAn}.  Here $p={\rm min} [1,x^{-\Delta h}]$ where $\Delta
h = \pm 1$ is the vertical displacement in the attempted elementary
move~\cite{nota1}, and $x = \exp(-mg/\kB T)$. We set $mg/\kB=1$ and 
$\nu=4$ throughout.
At high enough packing density, dynamical models of this kind possess
an extensive entropy of blocked states (defined as configurations in
which any particle is unable to move) whose derivative is the
so-called Edwards compactivity.
For this reason such models exhibit a slow compaction dynamics
reminiscent of dense granular matter~\cite{SeAr,LeArSe}.  It was found
in particular that during compaction a generalized
fluctuation-dissipation relationship is obeyed~\cite{Se2}, giving a
first evidence of an effective temperature in this regime.
In Ref.~\cite{Se2} the drift contribution to the longitudinal
mean-square displacement was ignored ~\cite{Jef}, leading to claim
that ``all measures of vertical correlation and response lead to the
impossibility of defining effective temperature'' and that ``the
vertical drift due to compaction leads to contradictory
results''~\cite{BaCoLo}.  Here we show that there are no such
contradictory results: in the slow compaction regime the drift brings
no qualitative change in the generalized fluctuation-dissipation
relation found in Ref.~\cite{Se2}, and its effect is substantially
negligible at high packing density.

The fluctuation-dissipation properties can be characterized by
applying a random perturbation to the system at times $t \ge \tw$:
\begin{eqnarray}
	{\cal H}_{\epsilon} &=& {\cal H}_0 + \epsilon \, \Theta(t-\tw)
	                        \sum_{i=1}^N f_i \, h_i \,\,,
\label{H}
\end{eqnarray}
where $f_i=\pm 1$ independently for each particle, $\epsilon$ is small
enough to probe the linear response regime, and $\Theta$ is the step
function. The integrated response function is then defined as:
\begin{eqnarray}
	\chi(t, \tw) &=& \frac{1}{N} \sum_{i=1}^N \left\langle \,
	\overline{f_i \Delta h_i(t) } \, \right\rangle \,,
\label{chi}
\end{eqnarray}
where $\Delta h_i(t)$ is the height difference between the perturbed
and unperturbed $i$ particle at time $t$. The angular brackets denote
the average over the thermal noise while the overline denotes
the average over the random force.
The `mean-square displacement' between two configurations at time
$\tw$ and $t > \tw$ is:
\begin{eqnarray}
	B(t,\tw) = \frac{1}{N} \sum_{i=1}^N \left\langle
	\left[h_i(t)-h_i(\tw)+\tilde h(\tw)-\tilde h(t) \right]^2
	\right\rangle ,
\label{B}
\end{eqnarray}
where the drift motion is taken into account by the average height:
\begin{eqnarray}	
	\tilde h(t) = \frac{1}{N} \sum_{i=1}^N \left\langle h_i(t)
	\right\rangle \,.
\label{h}
\end{eqnarray}
Fig.~\ref{X_tw13} reports a parametric plot of mean-square
displacement vs. response function at waiting time $\tw=2^{13}$: it
clearly shows that 
the presence of a slow longitudinal drift does not prevent the
existence of generalized fluctuation-dissipation relation. In
particular,
no qualitative change occurs in the characteristic
broken-line pattern when the drift term $\tilde h(\tw)-\tilde h(t)$ in
Eq.~(\ref{B}) is neglected, while appreciable quantitative deviations
between the two sets of data (with and without the drift term) only
appear when the measurement time is quite long.  Similar results were
also found in the so-called FILG model under gravity~\cite{VaAr}.
%
%
\begin{figure}
\begin{center}
\epsfig{file=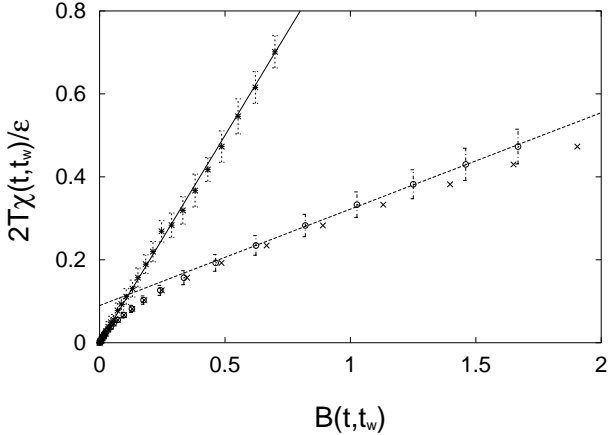,width=8.5cm}
\end{center}
\caption{ 
	Parametric plot of mean-square displacement $B(t,\tw)$ vs.
	response function $2 T \chi(t,\tw)/\epsilon$, during compaction
	dynamics (circle symbols).  The system is prepared in a random loose
	packed state with average density $\rho_{\rm rlp} \simeq 0.707$, and
	evolves under gravity and thermal vibration with $x=\exp(-1/T)=0.2$.  
	The perturbation is turned on at the waiting time $\tw = 2^{13}$ and 
	measurements are carried out for times $t$ in the range 
	$[\tw,\,\tw+10^5]$.  The slope of the dashed line is $0.23$ (to be
	compared with $0.20$ obtained by neglecting the drift term ($\times$
	symbols)).  The solid line with slope one is the equilibrium
	fluctuation-dissipation theorem, which is recovered by removing
	kinetic constraints (star symbols).  }
\label{X_tw13}
\end{figure}    
%
\begin{figure}
\begin{center}
\epsfig{file=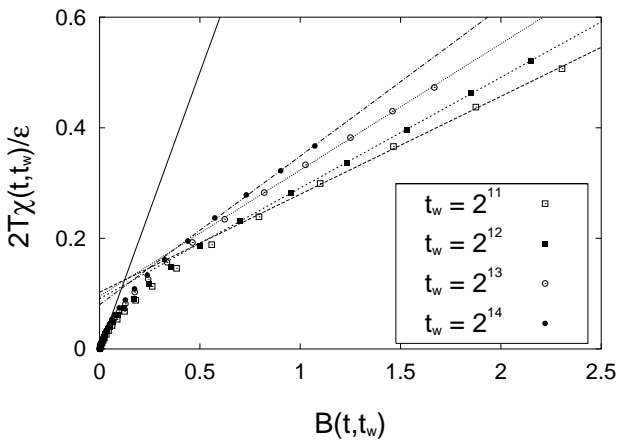,width=8.5cm}
\end{center}
\caption{
	Non-equilibrium fluctuation-dissipation relation in a
	compaction experiment as in FIG.~1, at different waiting 
	times $\tw$. The slope of the straight lines is 0.18, 0.20, 
	0.23, and 0.27, for increasing waiting time.	
	}
\label{X}
\end{figure}    
%
%
The second result of the numerical compaction experiment is reported
in Fig.~\ref{X}: it shows that the generalized fluctuation-dissipation
relation is obeyed at different waiting times.  From the parametric
plot of $\chi(t,\tw)$ vs. $B(t,\tw)$ one can define a time scale
dependent effective temperature by means of the relation:
\begin{eqnarray}
	\Tdyn (t,\tw) &=& \frac{\epsilon}{2}
	\frac{B(t,\tw)}{\chi(t,\tw)} \,, 
\label{T_dyn}
\end{eqnarray}
provided $\Tdyn$ is constant on that time scale, and
where now it is understood a possible dependence of $\Tdyn$ on
the density profile. Indeed, during compaction the system develops
inhomogeneous density profiles~\cite{SeAr,LeArSe}, as it also happens
after a sudden compression in zero gravity~\cite{PeSe}, and one may
wonder about their influence on the effective temperature.
In Ref.~\cite{LeArSe} the {\it stationary} density profile was
interpreted as formed by two parts: a lower flat part at critical
density $\rhoc \simeq 0.84$, and an upper equilibrium part in which
kinetic constraints play no role.  
Fig.~\ref{inset_prof_x02} shows the temporal evolution of the density
profile corresponding to the above compaction experiment: one observes
that the upper part of the bulk density profile increases faster than
the lower one (even when the former become denser than the latter),
and that the bulk profile is far from being flat (see inset of
Fig.~\ref{inset_prof_x02}).  At late time the contribution of the top
free interface is small for weak vibration, and - if sizeable - it
would make higher the slope $T/\Tdyn$, of the
fluctuation-dissipation plot (i.e. smaller the effective
temperature). While the contribution of particles at the bottom is
negligible as they do not evolve at all.
%
\begin{figure}
\begin{center}
\epsfig{file=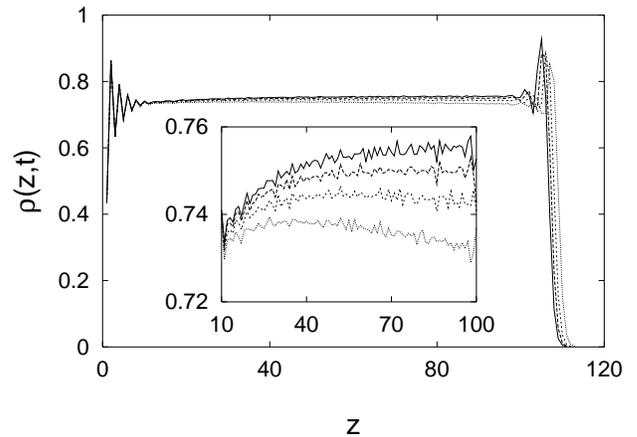,width=8.5cm}
\end{center}
\caption{
	Temporal evolution of the density profile during compaction
	dynamics ($x=0.2$, and time $t=2^{10+k}$ for $k=1$ to $4$). Inset:
	bulk density profile.}
\label{inset_prof_x02}
\end{figure}    
%
That the generalized fluctuation-dissipation relation is not affected
{\it qualitatively} by an inhomogeneous density profile can be
understood in terms of the mean-field dynamical model introduced
in~\cite{PeSe}, and further generalized to nonzero gravity
in~\cite{LeArSe}.  In both cases the long-time relaxation of the local
density factorizes: $\rhoc - \rho(z,t) = f(z)\, g(t)$. Since the mean
square displacement can be written as $B(z,t,\tw) = F(z) \, G(t,\tw)$
on long enough time interval $t-\tw$, the violation factor entails two
independent contributions: a purely geometric factor and a purely
dynamic one.  The latter contribution is only responsible for the
intrinsic violation of the fluctuation-dissipation relation. The
notion of effective temperature therefore seems to be still reliable
provided a geometric factor is taken into account (in this specific
case the global geometric factor entering Eq.~(\ref{T_dyn}) would be
${\cal F}=\int f(z) {\rm d}z/\int F(z) {\rm d}z$).  Notice that for a
purely flat density profile there is no difference between
`horizontal' and `vertical' observables, (and ${\cal F}=1$).

The question that naturally arises is whether the longitudinal effective
temperature can be interpreted in terms of the Edwards measure, which
should now be obtained {\it by fixing the density profile of the
experimental situation one wishes to reproduce}~\cite{nota2}.  The
numerical implementation of this strategy is however not
straightforward.  A more pragmatic approach consists in fixing a few
feature of the profile (as suggested in this case by the expression of
${\cal F}$), such as average density, average slope and so on. This is
quite similar in spirit to the construction of restricted Edwards measure
which has been recently exploited in Ref.~\cite{BeFrSe,Le}, and
generally improves the comparison with numerical experiments.
%
\begin{figure}
\begin{center}
\epsfig{file=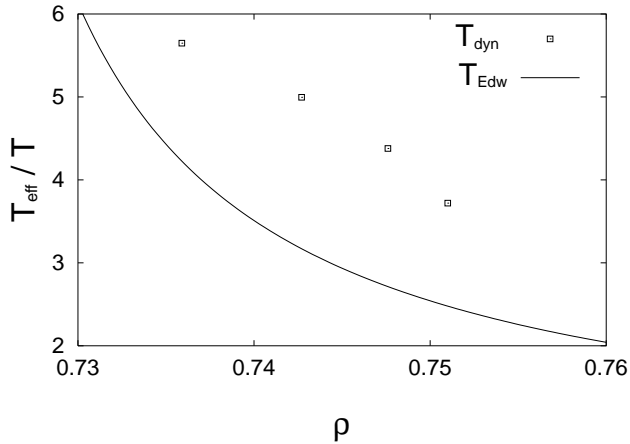,width=8.5cm}
\end{center}
\caption{ 
	Comparison between effective temperatures: $\Tdyn$ is
	measured during the compaction experiment from the
	fluctuation-dissipation relation at time $\tw$. $\TEdw$ is
	approximated through the Edwards measure with a homogeneous density
	$\rho$ such that the average bulk density profile at time $\tw$ of the
	compaction experiment is $\rho_{\rm av} (\tw)=\rho$.}
\label{T_Edw}
\end{figure}    
%
%
As a preliminary attempt to relate the effective temperature $T_{\rm
dyn}$, to the inverse compactivity $\TEdw$, we have computed the
Edwards entropy $\sEdw(\rho)$ to the lowest approximation, i.e. just
by fixing a homogeneous density.  In passing, the definition of
blocked configuration in presence of gravity requires here some care:
if one assumes a flat profile, it is not clear why blocked
configurations should depend upon gravity, like in the definition
adopted in Ref.~\protect{\cite{BaCoLo}}.  We have therefore explicitly
checked that there is no dependence upon gravity for the bcc
lattice. This is not however the most general case: interestingly, we
found that for the simple cubic lattice, the anisotropy due to
gravity, (say along the direction 001), suppresses the first order
character of the phase transition present in the Edwards measure at
density below $\approx 0.7$,~\cite{silvio}.  We have then estimated
$\TEdw (\rho)$ from the relation:
\begin{eqnarray}
	\TEdw\, \frac{{\rm d} \sEdw}{{\rm d} \rho} &=& T \, \frac{{\rm
	d} s}{{\rm d} \rho} \,,
\label{Eq:T_Edw}
\end{eqnarray}
where $s(\rho)= -\rho \log \rho - (1-\rho) \log (1-\rho)$ is the
equilibrium entropy~\cite{nota3}.  The two effective temperatures,
$\TEdw$ and $\Tdyn$, are shown in Fig.~\ref{T_Edw} at several
densities corresponding to the average bulk density profiles of the
compaction experiment.  It is clear that a ponderable comparison 
is possible only when other features of the density profile (e.g. the
average slope) are fixed, but much work is needed to test this point.

In conclusion, we confirm the existence of a generalized
fluctuation-dissipation relation in an abstract model of dense
granular matter which exhibits non-stationary inhomogeneous density
profiles.  The occurrence of a longitudinal effective temperature in
the slow compaction regime has been justified by a mean-field
dynamical model and its relation with the Edwards compactivity has
been discussed.
 
\bigskip

J.J. Arenzon, A. Barrat and S. Franz are acknowledged for discussions and
comments.



\begin{thebibliography}{99}
\bibitem{nota1} 
	Notice that there is no purely horizontal motion in a bcc
	lattice: any elementary (one lattice spacing) displacement is
	either upward or downward.
\bibitem{nota2} 
	This is indeed suggested by the functional nature of the
	nonzero gravity lattice gas entropy. For example, in the
	non-interacting case, Hamiltonian Eq.~(1), the equilibrium
	entropy is
	$$
	s[\rho] = -\sum_z\left\{ \rho(z) \log \rho(z) +
	\left[1-\rho(z) \right] \log \left[1-\rho(z) \right] \right\}
	\,.
	$$
	From the minimization of the free energy one then gets 
	$$\frac{mgz}{T} = \frac{\delta s}{\delta \rho(z)} \,.$$
\bibitem{nota3} 
	Notice that in the inhomogeneous case -- when the entropy is a
	functional of the density profile, the functional version of
	Eq.~(7) holds, and $\TEdw$ will only depend on the density
	profile provided its thermodynamic interpretation holds true.
\end{thebibliography}
\end{document}